\documentclass{article}
\usepackage{amssymb}
\usepackage{amsfonts}
\usepackage{amsmath}
\usepackage{epsfig}
\newcommand{\be}{\begin{eqnarray}}
\newcommand{\ee}{\end{eqnarray}}
\newcommand{\etal}{{\it et al.}}

\newcommand{\ms}{\Delta m^2_{21}}
\newcommand{\ma}{\Delta m^2_{31}}

\newcommand{\sss}{\sin^2 \theta_{12}}
\newcommand{\sch}{\sin^2 \theta_{13}}

\newcommand{\sig}{$3\sigma$}

\begin{document}
\title{%\hspace{4.1in}\\
%\hspace{4.1in}{\small OUTP XXXP}\\
\begin{flushright}
\small{OUTP-08 03 P}\\
\small{CERN-PH-TH/2008-015}\\
\end{flushright}
\bigskip
The effect of primordial fluctuations on neutrino oscillations}
%Probing neutrino oscillations from supernovae shock waves via a
%Megaton water Cerenkov detector.}
\author{N. P. Harries\thanks{email: \tt n.harries1@physics.ox.ac.uk}\\\\
{\normalsize \it The Rudolf Peierls Centre for Theoretical Physics,}\\
{\normalsize \it University of Oxford, 1 Keble Road, Oxford, OX13NP,
UK}\\
{\normalsize and}\\
{\normalsize \it TH Division, CERN, 1211 Geneva 23, Switzerland}}
\date{}
\maketitle
\begin{abstract}
Recent work has shown that neutrino oscillations in matter can be
greatly enhanced by flips between mass eigenstates if the medium is
fluctuating with a period equal to the neutrino oscillation length.
Here we investigate the effect of the primordial fluctuations on the
neutrino oscillations in the early universe. We calculate the
oscillation probability in the case of a general power law
fluctuation spectrum and for a more realistic spectrum predicted by
inflation. We also include the effect of the amplification of
fluctuations resulting from the QCD phase transition. We find that
there is a region of parameter space where this mechanism would be
the dominant mechanism for producing sterile neutrinos. However this
conclusion does not take account of the damping of fluctuations on
the neutrino oscillation scale when the neutrinos decouple from the
plasma. We find that this reduces the probability of flips between
the mass eigenstates to an unobservable level.
\end{abstract}
%
%%%%%%%%%%%%%%%%%%%%%%%%%%%%%%%%%%%%%%%%%%%%%%%%%%%%%%%%%%%%%%%%%%%%%%%%%%%%%%%%%%%%%%%
\section{Introduction}
%%%%%%%%%%%%%%%%%%%%%%%%%%%%%%%%%%%%%%%%%%%%%%%%%%%%%%%%%%%%%%%%%%%%%%%%%%%%%%%%%%%%%%%
%
For many years there has been tremendous success in determining the
masses and mixing of the neutrino sector, including experiments on
solar \cite{solar}, reactor \cite{kl2,chooz}, atmospheric
\cite{skatm} and accelerator \cite{k2k,minos} neutrinos. At the
\sig{} level the masses and mixing angles determining the solar
neutrino oscillations are $\ms=(7.2-9.2) \times 10^{-5}$ eV$^2$ and
$\sss=0.25-0.39$ \cite{solglobal}, while those determining the
atmospheric oscillations are $\ma=2.0-3.2 \times 10^{-3}$ and
$0.34<sin^2\theta_{23}<0.68$ with $\sin^22\theta_{23} < 0.9$
%\cite{atmglobal}
\cite{solglobal}. Currently the sign of $\Delta m^{2}_{23}$ is
undetermined and the current \sig{} upper bound on the allowed
values for the final mixing angle is $\sch < 0.044$
\cite{solglobal}. The LSND experiment showed oscillations on a mass
scale of $\Delta m^{2}\sim1eV^{2}$\cite{lsnd} if this observation is
interpreted as standard neutrino oscillations then it can only be
explained as oscillations into a forth neutrino state. Precision
measurements at LEP have shown that there are only three neutrinos
that interact via the standard weak interaction exist with mass less
than $M_{Z}/2$ and therefore these addition states must be sterile
\cite{sterileold}. The global data including the LSND (but not
including MiniBoone) comprehensively disfavored the inclusion of
just one sterile neutrino \cite{Maltoni:2002xd}, the inclusion of 2
sterile neutrinos mildly mixed with the 3 active neutrinos has a
more acceptable fit to all data including LSND \cite{threeplustwo}
although the value of the associated LSND mixing angle is still
problematic \cite{Strumia:2006db}. The inclusion of the MiniBoone
experiment did not observe an oscillation
\cite{Aguilar-Arevalo:2007it} but does not ruble out this scheme
\cite{Maltoni:2007zf}. Also $keV$ sterile neutrinos which naturally
arise in many models \cite{Brahmachari:2002va} can also explain
pulsar kicks \cite{Kusenko:1997sp} and are warm dark matter
candidates \cite{Bode:2000gq}. The latter has recently been received
interest following the observations of the central cores of low mass
galaxies \cite{Dalcanton:2000hn} and from the low number of
satellites observed in Milky-Way sizes galaxies
\cite{Kauffmann:1993gv}.

If sterile neutrinos do mix with the standard model neutrinos then
they would be produced via oscillations in the early universe
\cite{historical}. If the sterile neutrinos were produced before the
active neutrinos decouple $(T\gtrsim MeV)$ active neutrinos would
oscillate into sterile neutrinos, then the active neutrinos would be
repopulated, resulting in an increase in the number of relativistic
degrees of freedom. The expansion rate of the universe would
increase, which would lead to a higher freeze-out temperature and
therefore a higher neutron to proton ratio, leading to a higher
abundance of helium and other elements \cite{DiBari:2001ua}. This
increased expansion rate can also be probed by observations of the
CMB \cite{Hannestad:2005ex, Bell:2005dr}. These sterile neutrinos
would also contribute to the warm dark matter content of the
universe which would suppress the formation of large scale structure
\cite{Bell:2005dr,Hannestad:2004qu}.

Neutrino oscillations in the early unverse have been extensively
investigated and bounds have been placed on the mixing angles and
difference in mass squared \cite{historical}. However, these studies
have neglected the fact that there are temperature fluctuations in
the early universe. It has been shown that matter fluctuations can
cause level crossing between mass eigenstates
\cite{Friedland:2006ta}, resulting in an amplification of these
oscillations. In the case of active-sterile mixing this could lead
to an amplification in the production of sterile neutrinos. If the
spectrum of the fluctuations was known and the fluctuations were
large enough then new bounds could be placed on the masses and
mixing of the neutrino sector. Alternatively if the masses and
mixing was known then then constraints could be placed on the
spectrum of the fluctuations and on the models which predict them.

This paper is organized as follows. We begin by calculating the
level crossing probability of a two neutrino system in an expanding
universe with a general spectrum of temperature fluctuations. In
sections 3 and 4 we calculate the level crossing probability for the
case of a power law spectrum and a spectrum predicted by inflation
respectively. In section 5 we investigate the effect of neutrino
decoupling and in section 6 the effect of the QCD phase transition.
Finally we conclude in section 7.
%
%%%%%%%%%%%%%%%%%%%%%%%%%%%%%%%%%%%%%%%%%%%%%%%%%%%%%%%%%%%%%%%%%%%%%%%%%%%%%%%%%%%%%%%
\section{Neutrino evolution}
%%%%%%%%%%%%%%%%%%%%%%%%%%%%%%%%%%%%%%%%%%%%%%%%%%%%%%%%%%%%%%%%%%%%%%%%%%%%%%%%%%%%%%%
To consider the Neutrino oscillations in a fluctuating media we
follow the analysis developed in \cite{Friedland:2006ta}. The
evolution of a neutrino system is determined by the Schr\"{o}dinger
equation, $\textit{i}\frac{\partial}{\partial t}\nu=H^{0}(t)\nu$
\footnote{Here we use the notation $H^{0}$ for a hamiltonian in a
medium which is not fluctuating.}, where t is time, here $H^{0}$ and
$\nu$ are the Hamiltonian and neutrino wavefunction in flavour
basis. This can be rotated such that the hamiltonian is
instantaneously diagonalised\footnote{where $U^{\dagger} H^{0}U$ is
diagonal}, $\textit{i}\frac{\partial}{\partial
t}\nu_{m}^{0}=H_{m}^{0}\nu_{m}^{0}$, where $\nu_{m}^{0}=U\nu$ are
the instantaneous mass eigenstates. For a two neutrino system
\be
U(t)=\left(
           \begin{array}{cc}
             \cos\theta_{M} & \sin\theta_{M} \\
             -\sin\theta_{M} & \cos\theta_{M} \\
           \end{array}
\right), \ee
\be
H_{m}(t)=\left(
               \begin{array}{cc}
                 -\Delta_{m} & -\textit{i}\,d\theta_{m}/dx \\
                 \textit{i}\,d\theta_{m}/dx & \Delta_{m} \\
               \end{array}
 \right)\label{eq:HamMatt},
\ee
where $\Delta_{m}=\sqrt{\left(\Delta m^{2}\cos2\theta -
A\right)^2+\left(\Delta m^{2}\sin^{2}2\theta\right)^{2}}/4E$,
$\tan2\theta_{m}=\delta m^{2}\sin2\theta/(\Delta
m^{2}\cos2\theta-A)$, $\theta$ is the vacuum mixing angle, $\Delta
m^{2}$ is the difference in mass squared, $E$ is the energy of the
neutrino and $A$ is the difference in matter potentials of the two
neutrino system. Before the neutrinos have decoupled, for
temperatures, $T\gtrsim MeV$ the matter potential takes the form
\begin{eqnarray}
A&\simeq&-\Delta C_{v}G_{F}^{2}E^{2}T^{4}/\alpha \label{eq:MatterPotential}\\
&\simeq&-1.86\times 10^{-20}\Delta
C_{v}\left(\frac{E}{MeV}\right)^{2}\left(\frac{T}{MeV}\right)^4MeV^{2},
\end{eqnarray}
where $\alpha$ is the fine structure constant and
$C_{v}=14\pi\left(a-\sin^2\theta_{W}\right)\sin^2\theta_{W}/45$,
$a=3$ for $\nu_{e}$ and $\bar{\nu}_{e}$, $a=1$ for $\nu_{\mu}$,
$\bar{\nu}_{\mu}$, $\nu_{\tau}$ and $\bar{\nu}_{\tau}$ and $C_{v}=0$
for sterile neutrinos. We have assumed that the neutrino,
anti-neutrino asymmetry is of the same order as the baryon asymmetry
and therefore the contribution to the matter potential can be
neglected. There is a resonance when the neutrino system is
maximally mixed, when $A=\Delta m^{2}\cos2\theta$. As the matter
potential is negative for both neutrinos and anti-neutrinos there
will only be a resonance if the difference in mass squared is
negative. It is convenient to consider the hamiltonian as the sum of
the averaged Hamiltonian, $H^{0}$ and the fluctuated Hamiltonian,
$\delta H$, where the true hamiltonian, $H\equiv H^{0}+\delta H$ and
in the absence of fluctuations $\delta H=0$. The Schrodinger
equation is rotated to diagonalize the averaged Hamiltonian where
the fluctuations are teated as a perturbation. Now
$\textit{i}\frac{\partial}{\partial
t}\nu_{m}^{0}=H_{m}(t)\nu_{m}^{0}$, $H_{m}(t)=H_{m}^{0}+\delta
H_{m}$ where $H_{m}^{0}$ is given by Eq. (\ref{eq:HamMatt}) and
\be \delta H_{m}(t)=\frac{\delta V}{2}\left(
           \begin{array}{cc}
             -\cos2\theta_{m} & \sin2\theta_{m} \\
             \sin2\theta_{m} & \cos2\theta_{m}  \\
           \end{array}\label{eq:FlucHam}
         \right),
\ee
where $V=A/2E$ and $\delta V=V-V_{0}$, where $V_{0}=A_{0}/2E$ and
$A_{0}$ is the matter potential in the absence of fluctuations. If
the off diagonal elements are non-zero transitions between the mass
eigenstates can occur. If $\Delta_{m}\ll|d\theta_{m}/dt|$ the
neutrino propagation is non-adiabatic with or without fluctuations.
If $|d\theta_{m}/dt|\ll\Delta_{m}$ the evolution in the absence of
fluctuations is adiabatic and the level crossing between the mass
eigenstates is determined by the off diagonal elements in Eq
(\ref{eq:FlucHam}), which are $\delta V\sin2\theta/2$. In the
perturbative limit where the fluctuations are small the level
crossing probability is
\be P\simeq\left|\int_{t_i}^{t_{f}}dt\frac{\delta
V\sin2\theta_{m}}{2}\exp\left(\textit{i}\int^{t}dt'2\Delta_{m}\right)\right|^2.
\ee
Numerical simulations have shown that for large fluctuations the
system becomes depolarized, i.e. with the probability of detecting
each flavour of neutrino being 1/2; a rough criteria for
depolarization is the region $P\gtrsim1/2$. The neutrinos that we
consider are in thermal equilibrium with the plasma and therefore
have an energy spectrum which is approximately Fermi-Dirac and is
charectorised by the temperature of the plasma. For this spectrum
the average energy of the neutrinos $\langle E\rangle=yT$, where
$y\simeq3.1514$. For the rest of this paper we consider the
oscillations of a neutrino with an energy equal to the average
energy of the ensemble of neutrinos, i.e. $E=yT$. To calculate the
fluctuated hamiltonian for a neutrino with this energy we use
Eq.(\ref{eq:MatterPotential}) $V=A/2E\propto E T^{4}$, in the case
of small fluctuations $\delta V = 4V\delta T/T+V\delta E/E$. If the
neutrino is in thermal equilibrium with the fluctuations $\delta
E/E=\delta T/T$ and therefore $\delta V = 5V\delta T/T$. If the
neutrino is not in thermal equilibrium with the fluctuations then
$\delta E/E\simeq0$, i.e. the energy of the neutrino remains
approximately constant across the scale of the fluctuation. As we
will see in the section \ref{sec:NeutrinoDamp} the neutrinos
decouple from the fluctuations at the scale which amplifies the
neutrino oscillations before the time at which the neutrinos
oscillate. Therefore it is a good approximation to use assume that
the neutrinos are not in thermal equilibrium with the fluctuations
and $\delta V = 4V\delta$. Expanding the fluctuations into their
Fourier components this now becomes
\be P=\int\frac{d^{3}\textbf{k}}{(2\pi)^{3}}\mathcal{P}(k)\left|
\int_{t_i}^{t_{f}}dt\,2V\sin2\theta_{m}
\exp\left(\textit{i}\int^{t}dt'\left(2\Delta_{m}-\frac{k_{z}}{a}\right)\right)\right|^2,\label{eq:generalP}
\ee
where $\textbf{k}$ is the comoving wavevector,
$\mathcal{P}(k)\equiv|\delta_{T}(k)|^{2}$, $\delta_{T}(k)$ is the
Fourier transform of the temperature fluctuation which we have
assumed to be constant in time, $a$ is the scale factor and $z$ is
defined to be the direction of neutrino propagation. In this
derivation we have approximated the neutrino to be massless,
therefore $ds^{2}\simeq dt^{2}-a(t)^2dx^2=0$, and
$|\Delta\textbf{x}|=\int dt/a$. Note that the integral in Eq.
(\ref{eq:generalP}) is oscillatory except when
$2\Delta_{m}=k_{z}/a$, i.e. the fluctuation length is equal to the
neutrino oscillation length. To solve the time integral we use the
stationary phase approximation. Here we define new variables
\begin{eqnarray}
Q(a, k_{z})&\equiv& q(a)-k_{z}/h,\\
q(a)&\equiv& \frac{2a}{h}\Delta_{m} = \mu \left(a^{2} +\lambda a^{-4}\right),\\
F(a)&\equiv&\frac{2Va\sin2\theta_{m}}{h}=2\mu\lambda\sin2\theta,\\
\mu&\equiv&\frac{\Delta m^{2}}{2y T_{0} h},\\
\lambda&\equiv&\frac{2 \Delta C_{v}G_{F}^{2}yT_{0}^6}{\alpha \Delta m^{2} \cos2\theta},\\
\end{eqnarray}
where for temperatures of interest the universe is in the radiation
dominated era, where $T=T_{0}/a$, $y=E/T$, the hubble parameter
$H\equiv\dot{a}/a=h/a^{2}$ and $T_{0}$, $y$ and $h$ are constant. We
define $|I|^{2}$ such that
$P=\int\frac{d^{3}\textbf{k}}{(2\pi)^{3}}\mathcal{P}(k)\left|I\right|^2$,
applying the stationary phase approximation
\begin{eqnarray}
|I|^{2} & = & \left|\int da F(a)e^{\textit{i}\int Q(a',k_{z})da'}\right|^{2} \label{eq:ExactInt} \\
& \simeq& \left|
F(a_{s1}(k_{z}))e^{\textit{i}Q(a_{s1}(k_{z}),k_{z})}\sqrt{\frac{2\pi}{\textit{i}Q'(a_{s1}(k_{z}),k_{z})}}
+ (a_{s1}\rightarrow a_{s2}) \right|^{2}\\
& \simeq & \frac{2\pi
F(a_{s1}(k_{z}))^{2}}{|Q'(a_{s1}(k_{z}),k_{z})|} + (a_{s1}\rightarrow a_{s2})\nonumber\\
&&+ \frac{2\pi
F(a_{s1}(k_{z}))F(a_{s2}(k_{z}))}{\left|Q'(a_{s1}(k_{z}),k_{z})Q'(a_{s2}(k_{z}),k_{z})\right|}
\sin\int_{a_{s1}(k_{z})}^{a_{s1}(k_{z})}Q(a,
k_{z})da.\label{eq:ApproxInt}
\end{eqnarray}
The integral in Eq. (\ref{eq:ExactInt}) is oscillatory except where
$Q(k,a)=0$, this defines $a_{si}$ where i=1, 2, such that
$Q(a_{si},k_{z}(a_{si}))=0$, note there are two real solutions to
this equation. Once $|I|^2$ is integrated with respect to $k_{z}$
the final term of Eq. (\ref{eq:ApproxInt}) is averaged to zero and
therefore it is no longer considered. Integrating $|I|^{2}$ with
respect to $k_{z}$ is done so by a change of variables,
$dk_{z}=\frac{dk_{z}}{da{si}}da_{si}=hQ'(a_{si})ds_{si}$. Now the
limits of integration are from $k_{z}=[-\infty,\infty]$, this
corresponds to $a_{s1}=[0, a_{0}]$ and $a_{s1}=[a_{0}, \infty]$,
where $a_{0}$. Therefore the integral can be simplified to
\be P=\frac{h}{2\pi}\int_{0}^{\infty}dk_{r}\int_{0}^{\infty}da\,
k_{r}F(a)^{2}\mathcal{P}(\sqrt{k_{r}^{2}+k_{z}(a)^{2}}),
\label{eq:ProbPreFluct} \ee
where we have introduced the cylindrical polar co-ordinates $(k_{r},
k_{z}, \theta)$. Note for $a_{s1}$ there is a minus sign from
$Q'(a_{s1}(k_{z}),k_{z})=-|Q'(a_{s1}(k_{z}),k_{z})|$ and also the
limits of integration, where $k_{min}=a_{0}$ and $k_{max}=0$. This
results in a positive term in Eq (\ref{eq:ProbPreFluct}). Now to
calculate the level crossing probability we need the spectrum of the
fluctuations.
%
%%%%%%%%%%%%%%%%%%%%%%%%%%%%%%%%%%%%%%%%%%%%%%%%%%%%%%%%%%%%%%%%%%%%%%%%%%%%%%%%%%%%%%%
\section{Power Law Spectrum}
\begin{figure}[h]
\begin{center}
\includegraphics[width=10cm]{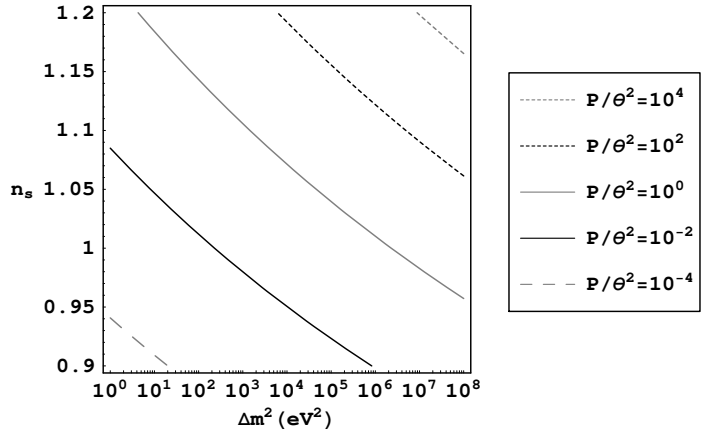}
\end{center}
\caption{The contours of constant $P/\theta^{2}$ as a function of
spectral index $n_{s}-1$ and difference in mass squared $\Delta
m^{2}$. \label{Fig:PowerSpec}}
\end{figure}

%%%%%%%%%%%%%%%%%%%%%%%%%%%%%%%%%%%%%%%%%%%%%%%%%%%%%%%%%%%%%%%%%%%%%%%%%%%%%%%%%%%%%%%
%
The measured spectrum of fluctuations is approximately flat. This
correlates to $\mathcal{P}=2\pi^{2}k^{-3}\Delta^{2}(k)$, where
$\Delta^{2}(k)=\Delta^{2}(k_{0})(k/k_{0})^{n_{s}-1}$, and the
spectral index $n_{s}-1$ is small. From the combination of the WMAP3
\cite{Hinshaw:2006ia} and SDSS \cite{Tegmark:2003uf} data,
$n_{s}=0.98\pm0.02$ at the $68\%$ confidence level
\cite{Peiris:2006sj}. To first approximation we assume that the
spectral index is constant. Our motivation for this is to analyze
the effect of how the parameters effect the level crossing
probability for the simplest case possible. For this case the level
crossing probability is
\be P\simeq\left[\Delta^{2}(k_{0})
\left(\frac{h\mu\cos2\theta\lambda^{1/3}}{k_{0}}\right)^{n_{s}-1}\right]
\left[\frac{\mu \lambda^{1/2}}{\cos2\theta}\right]\sin^{2}2\theta
\frac{4\pi}{2-n_{s}}\int_{0}^{u_{max}}\frac{u^{8-4(n_{s}-1)}}{(1+u^{6})^{3-(n_{s}-1)}}du,
\label{eq:PowerProb}\ee
where we have defined $u\equiv\lambda^{-1/6}a$. The first square
bracket is the size of the fluctuations at the neutrino oscillation
scale with a scaling factor of $\lambda^{1/3}$. From Eq.
(\ref{eq:generalP}) we can see that the fluctuations at the neutrino
oscillation length scale is the important factor in increasing the
level crossing probability. The second square bracket is the
weighted number of oscillation lengths. This can be more easily seen
if one expands $\mu\lambda^{1/2}=h^{-1}/(2yT_{0}/\Delta
m^{2}\lambda^{1/2})$ and realising that $h^{-1}$ is the scaled
Hubble length and $(2yT_{0}/\Delta m^{2}\lambda^{1/2})$ is the
scaled oscillation length. As one might naively expect the level
crossing probability increases with the number of oscillation
lengths. There is also a mixing factor $\sin^{2}2\theta$ and a
numerical factor which are as one would expect. Considering the
oscillation of electron neutrinos into sterile neutrinos the
oscillation probability is given by
\be P_{es}=1/2-(1/2-P)\cos2\theta_{i}\cos2\theta_{f}, \ee
where $\theta_{i}$ and $\theta_{f}$ are the mixing angles at
production and detection of the neutrino. For small mixing
$\cos2\theta\simeq1-2\theta^{2}$ and therefore
\be P_{es}\simeq(1-2P)(\theta_{i}^{2}+\theta_{f}^{2})+P.
\label{eq:Prob} \ee
For $P/\theta^{2}\gtrsim1$ the oscillation probability is dominated
by the level crossing probability caused by the fluctuating plasma
and is this dominant mechanism for the production of sterile
neutrinos. In Figure \ref{Fig:PowerSpec} we plot the contours of
constant $P/\theta^{2}$ as as function of mass squared differences
and spectral gradient. The level crossing probability increases with
$\Delta m^{2}$, this is because this corresponds to smaller
oscillation length and therefore more oscillation periods. Also the
level crossing probability increases with $n_{s}$, this is because
the amplitude of the fluctuations on the neutrino oscillation length
scale are larger.
%
%%%%%%%%%%%%%%%%%%%%%%%%%%%%%%%%%%%%%%%%%%%%%%%%%%%%%%%%%%%%%%%%%%%%%%%%%%%%%%%%%%%%%%%
\section{The primordial spectrum from Inflation}
\begin{figure}[h]
\begin{center}
\includegraphics[width=10cm]{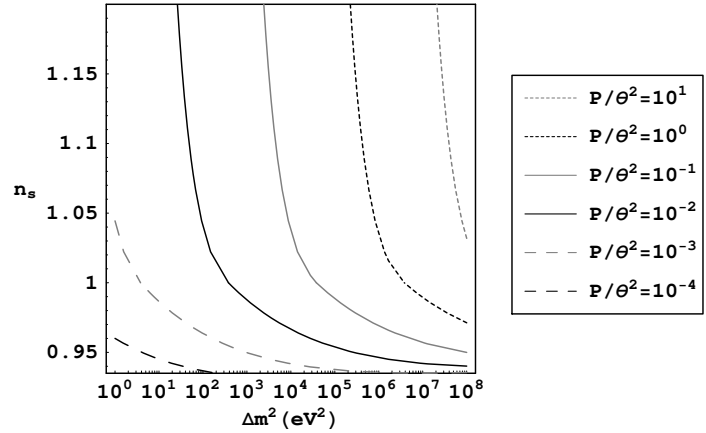}
\end{center}\label{fig:InflationSpectrum}
\caption{The contours of constant level crossing probability for a
spectrum predicted by inflation with a potential $V=a_{n}\phi^{n}$
where $n=-1$ for $n_{s}>1$ and $n=4$ for $n_{s}<1$.}
\end{figure}

%%%%%%%%%%%%%%%%%%%%%%%%%%%%%%%%%%%%%%%%%%%%%%%%%%%%%%%%%%%%%%%%%%%%%%%%%%%%%%%%%%%%%%%
%
An initial stage of inflation seems a necessary ingredient ot the
Big Bang model\cite{Guth:1980zm}. An intriguing feature of
inflationary models is that they predict a nearly scale invariant
spectrum which is consistent with that observed by WMAP
\cite{Hinshaw:2006ia}, the 2dFGRS \cite{Cole:2005sx} and SDSS
\cite{Tegmark:2003uf}. A simple model capable of generating this
spectrum of fluctuations for inflation is a single scaler field,
$\phi$, with the potential $V(\phi)$, in the slow role approximation
$\varepsilon\equiv (M_{pl}V'/V)^{2}/2\ll1$ and $|\eta|\equiv
M_{pl}^2|V''/V|\ll1$, where $'$ represents $d/d\phi$. For this model
the curvature fluctuations, $\Delta_{R}^{2}(k)$, which are related
to the temperature fluctuations by
$\mathcal{P}(k)=2\pi^{2}k^{-3}\Delta_{R}^{2}(k)$ are given by
\be
\Delta_{R}^{2}(k)=\left.\frac{H^4}{(2\pi\dot{\phi})^{2}}\right|_{t_{*}},
\ee
where $t_{*}$ is the time at which the $k$ mode crossed the horizon,
$k=aH|_{t_{*}}$. This together with the equation of motion for the
inflaton field in the slow role approximation, $3H\dot{\phi}=-V'$,
and the Hubble parameter, $H^{2}=8\pi V/3M_{pl}^{2}$ results in
\be \Delta_{R}^{2}(k)=\left.\frac{128\pi}{9
M_{pl}^2}\frac{V^{3}}{V'^{2}}\right|_{t_{*}}. \ee
Approximating $H$ to be constant and using $k=aH|_{t_{*}}$
\be \ln(k/k_{0})=-\int_{\phi_{0}}^{\phi}\frac{8 \pi V
d\phi}{M_{pl}^{2}V'}. \ee
For simplicity we consider the potential $V=a_{n}\phi^{n}$, for this
case
\be
\Delta_{R}^{2}(k)=\frac{128}{3n^{2}}\frac{V_{0}}{M_{pl}^{4}}\left(\frac{\phi_{0}}{M_{pl}}\right)^{2}
\left(1-\frac{n
M_{pl}^{2}}{4\pi\phi_{0}^{2}}\ln(k/k_{0})\right)^{(2+n)/2}.
\label{eq:InfInSp}\ee
where $V_{0}=a_{n}\phi_{0}^{n}$. We can see that the spectrum is
flat for $n=0,-2$, i.e. for $n_{s}=1$, increases with k for
$-2<k<0$, for $n_{s}>1$ and decreases with k for $n>0$ and $n<-2$,
for $n_{s}<1$. Now to obtain analytical results we can use
observations of the primordial spectrum to determine the three of
the four free parameters. At the $68\%$ confidence level
$\ln\left(\Delta_{R}^{2}(k_{0})10^{10}\right)=3.17\pm0.06$ and
$d\ln\Delta_{R}^{2}/d\ln k |_{k_{0}}=n_{s}(k_{0})-1=-0.02\pm0.02$,
where $k_{0}=0.002Mpc^{-1}$ \cite{Peiris:2006sj}, re-expressing Eq
(\ref{eq:InfInSp}) in terms of these parameters
\be
\Delta_{R}^{2}(k)=\Delta_{R}^{2}(k_{0})\left(1+\frac{n_{s}(k_{0})-1}{\alpha_{n}}
\ln\left(k/k_{0}\right)\right)^{\alpha_{n}}, \ee
where we have chosen $\alpha_{n}\equiv(2+n)/2$ to be
undetermined\footnote{For fluctuations to be present at the neutrino
oscillation length scale we require that the number of e-folds,
$N_{e}=\int da/a\simeq60\lesssim\ln(k_{0}/k)$, for mass squared
differences up to $keV^{2}$ this is true.}. \newline For this
spectrum of primordial fluctuations the level crossing probability
is
\begin{eqnarray}
P&=&2\pi\Delta_{R}^{2}\mu\lambda^{1/2}\sin^{2}2\theta \int_{0}^{\infty}dv\int_{0}^{u_{m}}du\,I_{2} \label{eq:InfProb}\\
I_{2}&=&\frac{u^{4}\left(v+\left(u^{2}+u^{-4}\right)^{2}\right)^{-3/2}}{(1+u^{6})^{2}}
\left(1-\frac{1-n_{s}(k_{0})}{\alpha_{n}}
\left(\ln\left(v+\left(u^{2}+u^{-4}\right)^{2}\right)
+\ln(h\mu\lambda^{1/3}/k_{0})\right)\right)^{\alpha_{n}}
\end{eqnarray}
Figure \ref{fig:InflationSpectrum} plots the contours of constant
$P/\theta^{2}$ for $n=-1$ for $n_{s}>1$ and $n=4$ for $n_{s}<1$. The
fluctuations increase with $\Delta m^{2}$ and $n_{s}$ for as for the
case of the power law spectrum. However the increase with $n_{s}$ is
not as pronounced as there is only a logarithmic increase with
$n_{s}$. This mechanism becomes dominant in the production of
sterile neutrinos for $keV$ neutrinos with $n_{s}>1$.
%
%%%%%%%%%%%%%%%%%%%%%%%%%%%%%%%%%%%%%%%%%%%%%%%%%%%%%%%%%%%%%%%%%%%%%%%%%%%%%%%%%%%%%%%
\section{Neutrino Damping}\label{sec:NeutrinoDamp}
\begin{figure}[h]
\begin{center}
\includegraphics[width=10cm]{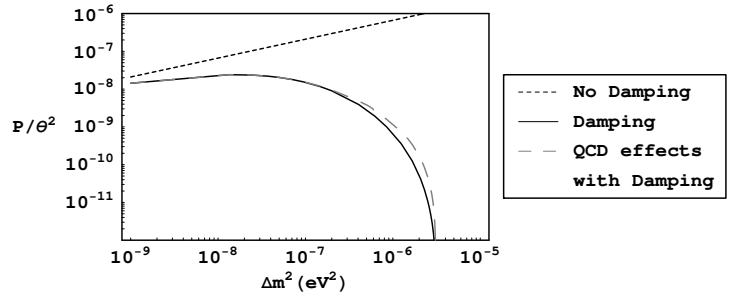}
\end{center}
\caption{The effects of damping on the power law spectrum with and
without fluctuation caused by the QCD phase transition.
\label{Fig:DampPowSpec}}
\end{figure}

%%%%%%%%%%%%%%%%%%%%%%%%%%%%%%%%%%%%%%%%%%%%%%%%%%%%%%%%%%%%%%%%%%%%%%%%%%%%%%%%%%%%%%%
%
So far we have neglected the damping of the fluctuations in the
early universe by neutrino diffusion which is equivalent to Silk
damping of photon decoupling. This occurs when the neutrinos diffuse
from over dense regions dragging the matter with them. Neutrinos
decouple at $T_{\nu_{e}}^{dec}\simeq1.4MeV$ for $\nu_{e}$ and
$T_{\nu_{\mu}\nu_{\tau}}^{dec}\simeq2.2MeV$ for $\nu_{\mu}$ and
$\nu_{\tau}$ \cite{Heckler:1993nc}, this occurs when universe
expands faster than the neutrino interaction rate $H>\Gamma_{\nu}$,
where $\Gamma_{\nu}$ is the neutrino interaction rate. However, the
neutrinos decouple from a particular mode of oscillation when
$c_{s}k_{ph}>\Gamma_{\nu}$, where $c_{s}$ is the sound speed of the
oscillation. Following \cite{Schwarz:2003du} the damping factor in
the density fluctuations is
\be D(k, \eta)\equiv\exp\left(-\Gamma\left(k,
\eta\right)\right)=\exp\left(-\frac{1}{2}\int_{0}^{\eta_{max}}
(k_{phys}/\epsilon_{tot})\eta_{visc}kd\eta\right)
\label{eq:DampSol}\ee
where $\rho_{tot}$ is the total energy density,
$\eta_{visc}=(4/15)\Sigma\rho_{\nu_{\alpha}}\tau_{\nu_{\alpha}}$ is
the shear viscosity co-efficient, $\rho_{\nu_{\alpha}}$ is the
energy density of $\nu_{\alpha}$,
$\tau_{\nu_{\alpha}}\equiv1/\Gamma_{\nu_{\alpha}}$ is the typical
collision time and $\Gamma_{\nu_{\alpha}}$ is the interaction rate
of $\nu_{\alpha}$ with the medium,
$\eta_{max}=min[\eta,\eta_{dec}(k)]$, $\eta_{dec}(k)$ is the
comoving time at the decoupling of the oscillation mode $k$ and the
sum is over all active neutrino species. Substituting the neutrino
interaction rate $\Gamma_{\nu}=\gamma_{\nu}G_{F}^{2}T^{5}$, the
energy density $\epsilon_{\alpha}=\pi^{2}g_{\alpha}(T) T^{4}/30$,
where $g$ is the effective number of relativistic helicity degrees
of freedom\footnote{where fermionic degrees of freedom are
suppressed by a factor of $7/8$} into Eq. (\ref{eq:DampSol})
\be \Gamma(k,T)=\sum_{\nu}\frac{4}{75}\frac{g_{\nu}}{g_{tot}}
\left(\frac{H}{\Gamma_{\nu_{\alpha}}}\right)_{T_{max}}
\left(\frac{k}{aH}\right)^{2}_{T_{max}},\label{eq:DampExp}\ee
where $T_{max}=max[T,T^{dec}_{\nu}(k)]$, $T^{dec}_{\nu}(k)$
decoupling temperature of $\nu$ with the oscillation mode $k$. The
decoupling temperature of an oscillation mode can be related to the
neutrino decoupling temperature by
\be T^{dec}_{\nu}(k)=T^{dec}_{\nu}\left(\frac{c_{s}k}{a
H}\right)^{1/4}_{T^{dec}_{\nu}},\ee
where $T^{dec}_{\nu}$ is the $\nu$ decoupling temperature. For all
$k$ modes the neutrino has decoupled before the neutrino oscillation
length is equal to the oscillation length (see appendix A) and
therefore $T_{max}=T_{dec}(k)$ and the damping is maximal. This
damping modifies the primordial spectrum to
\be
\Delta^{2}(k)\rightarrow\Delta^{2}(k)\exp\left(-\Gamma\left(k\right)\right),\ee
\be
\Gamma\left(k\right)=\Lambda\left(\left(u^{2}+u^{-4}\right)^{2}+v\right)^{3/8},
\ee
where
\be \Lambda\equiv\sum_{\nu}\frac{4}{75}\frac{g_{\nu}}{g_{tot}}
\left(\frac{H}{\Gamma_{\nu_{\alpha}}}\right)_{T_{dec}}c_{s}^{-5/4}
\left(\mu_{dec}\lambda_{dec}^{1/3} \cos2\theta\right)^{3/4}, \ee
where $\lambda_{dec}=\lambda a_{dec}^{-6}$, $\mu_{dec}=\mu
a_{dec}^{3}$ and $a_{dec}$ is the scale factor at the time of
neutrino decoupling. This effectively gives a cutoff in the
fluctuation spectrum when
$\left(\left(u^{2}+u^{-4}\right)^{2}+v\right)=\Lambda^{-8/3}=3.5\times
10^{-10}\left(\frac{eV^{2}}{\Delta m^{2}}\right)^{4/3}$. As
$\Lambda\propto\Delta \left(m^{2}\right)^{-4/3}$ the cutoff is
larger for smaller mass differences. This is because smaller mass
differences corresponds to a larger oscillation length, therefore
fluctuations with smaller $k$. From Eq. (\ref{eq:DampExp}) we see
that damping is largest for (a)larger $k/\Gamma\sim\tau/L_{osc}$,
where $L_{osc}$ is the oscillation wavelength and (b)$k/H$.
Physically (a) is due to the neutrinos 'dragging' the matter to
distances larger than the oscillation wavelength and (b) is for
smaller expansion rate and larger oscillation frequency there will
be a larger number of oscillation lengths when the damping is large.
The effect of this damping on the power spectrum with $n_{s}=1$ is
shown in Figure \ref{Fig:DampPowSpec}. Here we see that the damping
severely reduces the level crossing probability for $\Delta
m^{2}\gtrsim 10^{-7}$. This damps any oscillations into sterile
neutrinos in the $eV-keV$ mass range and the oscillations between
active neutrinos.
%%%%%%%%%%%%%%%%%%%%%%%%%%%%%%%%%%%%%%%%%%%%%%%%%%%%%%%%%%%%%%%%%%%%%%%%%%%%%%%%%%%%%%%
\section{QCD phase transition}
%%%%%%%%%%%%%%%%%%%%%%%%%%%%%%%%%%%%%%%%%%%%%%%%%%%%%%%%%%%%%%%%%%%%%%%%%%%%%%%%%%%%%%%
%
At a temperature of about $T_{*}\sim150MeV$ there is a QCD phase
transition from a quark-gluon plasma to a hadron gas, this
transition can increase the amplitude of the primordial fluctuations
for sub horizon modes while leaving the superhorizon modes
unaffected \cite{Schmid:1996qd}. According to the second law of
thermodynamics
\be \rho + p = T \frac{dp}{dT} \ee
where $\rho$ and $p$ are the energy density and pressure
respectively. During the phase transition the energy density is
discontinuous in temperature at the temperature of the phase
transition $T_{*}$, therefore the pressure must be continuous with a
discontinuous gradient. During the phase transition the energy
density decreases from $\rho_{-}(T_{*})$ to $\rho_{+}(T_{*})$, where
we now use the notation $-$ and $+$ for the beginning and end of the
transition respectively, the pressure p remains constant $p(T_{*})$,
and therefore the speed of sound $c_{s}=\left(\partial
p/\partial\rho\right)_{s}^{1/2}=0$. As the sound speed is zero there
are no restoring forces from pressure gradients and therefore the
radiation fluid goes into free fall. This can be seen quantitatively
from the equations of motion for sub horizon density fluctuations
\begin{eqnarray}
\delta'-k\hat{\psi}=0\\
\hat{\psi}'+c_{s}^{2}\delta=0
\end{eqnarray}
where $\hat{\psi}=(\rho/(\rho+p))v_{pec}$ and $v_{pec}$ is the
peculiar velocity. During the phase transition $\hat{\psi}'=0$ and
$\delta'=k\hat{\psi}=const$. The solution to these equations are
$\hat{\psi}=\hat{\psi}_{-}$ and
$\delta=\delta_{-}+k(\eta-\eta_{-})\hat{\psi}_{-}$, i.e. the energy
density increases linearly with the time of the phase transition and
the wavevector. This results in the ratio of the initial
fluctuations, $A_{i}$, to the final oscillations, $A_{f}$ being
$A_{f}/A_{i} \simeq k/k_{1}$ where
$k_{1}=\sqrt{3}/(\eta_{+}-\eta_{-})$.
This approximately modifies the spectrum of density fluctuations by
\be \Delta^{2}(\eta, k)\rightarrow \left\{
    \begin{array}{cc}
    \Delta^{2}(\eta, k) & for k\leqslant k_{1}\\
    \Delta^{2}(\eta, k)(k/k_{1})^2 & for k>k_{1}\\
    \end{array}
    \right.
\ee
$P/\theta^{2}$ is plotted in Figure \ref{Fig:DampPowSpec} including
the effects of the QCD phase transition. From this we see that the
level crossing probability is increased but the fluctuations are
still severely damped and consequently there is still no observable
effect.
%%%%%%%%%%%%%%%%%%%%%%%%%%%%%%%%%%%%%%%%%%%%%%%%%%%%%%%%%%%%%%%%%%%%%%%%%%%%%%%%%%%%%%%
\section{Conclusion}
%%%%%%%%%%%%%%%%%%%%%%%%%%%%%%%%%%%%%%%%%%%%%%%%%%%%%%%%%%%%%%%%%%%%%%%%%%%%%%%%%%%%%%%
%
If sterile neutrinos mix with the standard model neutrinos they will
be produced by oscillations in the early universe and the number
produced will depend on the mass squared differences and mixing
angles. Sterile neutrinos would add to the relativistic energy
density of the universe increasing the expansion rate of the
universe resulting in an increase in the abundance of light
elements. They would also contribute to the warm dark matter of the
universe. Neutrino oscillations propagating in a fluctuating medium
can be enhanced due to the increase in the level crossing
probability between mass eigenstates. This probability is dominated
by the amplitude of the fluctuations on the neutrino oscillation
scale. The temperature fluctuations in the the early universe have
been probed by the WMAP, 2dFGRS and SDSS data and have shown that
this spectrum of fluctuations is nearly scale invariant. In this
paper we have calculated the level crossing probability for the case
the spectrum of density fluctuations has a simple power law
dependence and for the case this is modified by logarithmic running
as predicted by a simple model of inflation. For these cases there
is a region of parameter space for large $\Delta m^{2}$ and $n_{s}$
were the production of sterile neutrinos would be dominated by this
mechanism. However this conclusion ignores the damping of the
fluctuation which occurs when the neutrinos decouple from the
plasma, this is equivalent to Silk damping. We have shown that the
neutrinos damp the fluctuations on the scale of the neutrino
oscillation length scale and thus the level crossing probability is
vastly reduced and is unobservable. A further effect is that the
fluctuations on the neutrino oscillation scale are amplified at the
QCD phase transition. However we find that this enhancement is
insufficient to overcome the damping that occurs. In conclusion the
production of sterile neutrinos does not get amplified from level
crossing between the mass eigenstates due to primordial fluctuations
because when the neutrinos decouple they damp the fluctuations at
the neutrino oscillation scale.
\section{Acknowledgements}
I would like to thank Graham Ross and Subir Sarkar for extremely
useful discussions. This work was supported by STFC Studentship
Award PPA/S/S/2004/03926 and by a Marie Curie Early Stage Research
Training Fellowship of the European Community's Sixth Framework
Programme under contract (MRTN-CT-2006-0355863-UniverseNet).
%
%appendix
%
\section*{Appendix A}
The decoupling temperature of the $k$ oscillation mode is
\be T^{dec}_{\nu}(k)=T^{dec}_{\nu}\left(\frac{c_{s}k}{a
H}\right)^{1/4}_{T^{dec}_{\nu}}.\ee
The wavevector of the the neutrinos oscillation is
\be
k=h\mu\lambda^{1/3}\cos2\theta\sqrt{v+\left(u^{2}+u^{-4}\right)^{2}}
\ee
where $u=\lambda^{-1/6}a=\lambda^{-1/6}T/T_{0}$ where $T$ is the
temperature of the universe. The ratio of the decoupling temperature
to the temperature of the universe is
\be
\frac{T^{dec}_{\nu}(k)}{T}=\left(c_{s}\mu_{dec}\lambda_{dec}\cos2\theta
\,u^{4}\sqrt{v+\left(u^{2}+u^{-4}\right)^{2}}\right)^{1/4}, \ee
where $\lambda_{dec}=\lambda a_{dec}^{-6}$, $\mu_{dec}=\mu
a_{dec}^{3}$ and $a_{dec}$ is the scale factor at neutrino
decoupling. The right hand side is minimized for $v=0$ and $u=0$,
where at this minimum
\be
\left.\frac{T^{dec}_{\nu}(k)}{T}\right|_{min}=\left((3/2^{2/3})c_{s}\mu_{dec}\lambda_{dec}\cos2\theta\right)^{1/4}
\simeq2.5.\ee
As the minimum of this ratio is greater than unity the temperature
of the universe when the wavelength of the fluctuation is equal to
the neutrino oscillation length is always greater than the
temperature at when the neutrinos decoupled from this mode.


\begin{thebibliography}{100}

\bibitem{solar}
B.~T.~Cleveland {\it et al.}, Astrophys.\ J.\  {\bf 496}, 505
(1998);
%
J.~N.~Abdurashitov {\it et al.}  [SAGE Collaboration], J.\ Exp.\
Theor.\ Phys.\  {\bf 95}, 181 (2002);
  [Zh.\ Eksp.\ Teor.\ Fiz.\  {\bf 122}, 211 (2002)]
%
W.~Hampel {\it et al.}  [GALLEX Collaboration], Phys.\ Lett.\ B {\bf
447}, 127 (1999);
%
C. Cattadori, Talk at Neutrino 2004, Paris, France, June 14-19,
2004;
%
S.~Fukuda {\it et al.}  [Super-Kamiokande Collaboration], Phys.\
Lett.\ B {\bf 539}, 179 (2002);
%
B.~Aharmim {\it et al.}  [SNO Collaboration], Phys.\ Rev.\ C {\bf
72}, 055502 (2005).

\bibitem{kl2}
T.~Araki {\it et al.}  [KamLAND Collaboration], Phys.\ Rev.\ Lett.\
{\bf 94}, 081801 (2005).

\bibitem{chooz}
M.~Apollonio {\it et al.}, Eur.\ Phys.\ J.\ C {\bf 27}, 331 (2003);
%
F.~Boehm {\it et al.}, Phys.\ Rev.\ D {\bf 64}, 112001 (2001).

\bibitem{skatm}
Y.~Ashie {\it et al.}  [Super-Kamiokande Collaboration], Phys.\
Rev.\ D {\bf 71}, 112005 (2005).

\bibitem{k2k}E.~Aliu {\it et al.}  [K2K Collaboration],
Phys.\ Rev.\ Lett.\  {\bf 94}, 081802 (2005).

\bibitem{minos}
  D.~G.~Michael {\it et al.}  [MINOS Collaboration],
  %``Observation of muon neutrino disappearance with the MINOS detectors and
  %the NuMI neutrino beam,''
  Phys.\ Rev.\ Lett.\  {\bf 97}, 191801 (2006).
%  [arXiv:hep-ex/0607088].
  %%CITATION = HEP-EX 0607088;%

\bibitem{solglobal}
M.~Maltoni {\it et al.},
%T.~Schwetz, M.~A.~Tortola and J.~W.~F.~Valle,
%%``Status of global fits to neutrino oscillations,''
New J.\ Phys.\  {\bf 6}, 122 (2004), hep-ph/0405172 v5;
%%%CITATION = HEP-PH 0405172;%%
%
  S.~Choubey,
%   ``Probing The Neutrino Mass Matrix In Next Generation Neutrino Oscillation
  %Experiments,''
  arXiv:hep-ph/0509217;
%
  S.~Goswami,
  %``Neutrino oscillations and masses,''
  Int.\ J.\ Mod.\ Phys.\ A {\bf 21}, 1901 (2006);
%
  A.~Bandyopadhyay {\it et al.},
%S.~Choubey, S.~Goswami, S.~T.~Petcov and D.~P.~Roy,
%   ``Update Of The Solar Neutrino Oscillation Analysis With The 766-Ty  Kamland Spectrum,''
  Phys.\ Lett.\ B {\bf 608}, 115 (2005);
%  [arXiv:hep-ph/0406328].
%
  G.~L.~Fogli {\it et al.},
%, E.~Lisi, A.~Marrone and A.~Palazzo,
  %``Global analysis of three-flavour neutrino masses and mixings,''
  Prog.\ Part.\ Nucl.\ Phys.\  {\bf 57}, 742 (2006).
%  [arXiv:hep-ph/0506083].

\bibitem{lsnd}
C. Athanassopoulos \etal, (The LSND Collaboration) Phys. Rev. Lett.
{\bf 77}, 3082 (1996); C. Athanassopoulos \etal, (The LSND
Collaboration) Phys. Rev. Lett. {\bf 81}, 1774 (1998).

\bibitem{sterileold}
  J.~J.~Gomez-Cadenas and M.~C.~Gonzalez-Garcia,
  %``Future tau-neutrino oscillation experiments and present data,''
  Z.\ Phys.\ C {\bf 71}, 443 (1996);
%  [arXiv:hep-ph/9504246].
  %%CITATION = HEP-PH 9504246;%%
%
  S.~Goswami,
  %``Accelerator, reactor, solar and atmospheric neutrino oscillation: Beyond
  %three generations,''
  Phys.\ Rev.\ D {\bf 55}, 2931 (1997).
%  [arXiv:hep-ph/9507212].
  %%CITATION = HEP-PH 9507212;%%
%
  S.~M.~Bilenky, C.~Giunti and W.~Grimus,
  %``Neutrino mass spectrum from the results of neutrino oscillation
  %experiments,''
  Eur.\ Phys.\ J.\ C {\bf 1}, 247 (1998).
%  [arXiv:hep-ph/9607372].
  %%CITATION = HEP-PH 9607372;%%

\bibitem{Maltoni:2002xd}
  M.~Maltoni, T.~Schwetz, M.~A.~Tortola and J.~W.~F.~Valle,
  %``Ruling out four-neutrino oscillation interpretations of the LSND
  %anomaly?,''
  Nucl.\ Phys.\  B {\bf 643}, 321 (2002)
  %[arXiv:hep-ph/0207157].

\bibitem{threeplustwo}
  M.~Sorel, J.~M.~Conrad and M.~H.~Shaevitz,
  %``A combined analysis of short-baseline neutrino experiments in the (3+1)
  %and (3+2) sterile neutrino oscillation hypotheses,''
  Phys.\ Rev.\ D {\bf 70}, 073004 (2004).
%  [arXiv:hep-ph/0305255].
  %%CITATION = HEP-PH 0305255;%%

\bibitem{Strumia:2006db}
  A.~Strumia and F.~Vissani,
  %``Neutrino masses and mixings and.,''
  arXiv:hep-ph/0606054.
  %%CITATION = HEP-PH/0606054;%%


\bibitem{Aguilar-Arevalo:2007it}
  A.~A.~Aguilar-Arevalo {\it et al.}  [The MiniBooNE Collaboration],
  %``A search for electron neutrino appearance at the Delta(m**2) ~ 1-eV**2
  %scale,''
  arXiv:0704.1500 [hep-ex].
  %%CITATION = ARXIV:0704.1500;%%


\bibitem{Maltoni:2007zf}
  M.~Maltoni and T.~Schwetz,
  %``Sterile neutrino oscillations after first MiniBooNE results,''
  arXiv:0705.0107 [hep-ph].
  %%CITATION = ARXIV:0705.0107;%%

\bibitem{Brahmachari:2002va}
  B.~Brahmachari, S.~Choubey and R.N.~Mohapatra,
  %``A minimal three generation seesaw scenario for LSND,''
  Phys.\ Lett.\  B {\bf 536}, 94 (2002).
%  [arXiv:hep-ph/0204073].
  %%CITATION = PHLTA,B536,94;%%
  Z.G.~Berezhiani and R.N.~Mohapatra,
  %``Reconciling present neutrino puzzles: Sterile neutrinos as mirror
  %neutrinos,''
  Phys.\ Rev.\  D {\bf 52}, 6607 (1995).
%  [arXiv:hep-ph/9505385].
  %%CITATION = PHRVA,D52,6607;%%
  E.J.~Chun and H.B.~Kim,
  %``Nonthermal axino as cool dark matter in supersymmetric standard model
  %without R-parity,''
  Phys.\ Rev.\  D {\bf 60}, 095006 (1999).
%  [arXiv:hep-ph/9906392].
  %%CITATION = PHRVA,D60,095006;%%
  P.~Langacker,
  %``A mechanism for ordinary-sterile neutrino mixing,''
  Phys.\ Rev.\  D {\bf 58}, 093017 (1998).
%  [arXiv:hep-ph/9805281].
  %%CITATION = PHRVA,D58,093017;%%
  K.N.~Abazajian, G.M.~Fuller and M.~Patel,
  %``The cosmological bulk neutrino catastrophe,''
  Phys.\ Rev.\ Lett.\  {\bf 90}, 061301 (2003).
%  [arXiv:hep-ph/0011048].
  %%CITATION = PRLTA,90,061301;%%
  T.~Asaka, S.~Blanchet and M.~Shaposhnikov,
%  %``The nuMSM, dark matter and neutrino masses,''
  Phys.\ Lett.\  B {\bf 631}, 151 (2005).
%  [arXiv:hep-ph/0503065].
%  %%CITATION = PHLTA,B631,151;%%

\bibitem{Kusenko:1997sp}
  A.~Kusenko and G.~Segre,
  %``Neutral current induced neutrino oscillations in a supernova,''
  Phys.\ Lett.\  B {\bf 396}, 197 (1997).
%  [arXiv:hep-ph/9701311].
  %%CITATION = PHLTA,B396,197;%%
  A.~Kusenko and G.~Segre,
  %``Pulsar kicks from neutrino oscillations,''
  Phys.\ Rev.\  D {\bf 59}, 061302(R) (1999).
%  [arXiv:astro-ph/9811144].
  %%CITATION = PHRVA,D59,061302;%%
  G.M.~Fuller, A.~Kusenko, I.~Mocioiu and S.~Pascoli,
  %``Pulsar kicks from a dark-matter sterile neutrino,''
  Phys.\ Rev.\  D {\bf 68}, 103002 (2003).
%  [arXiv:astro-ph/0307267].
  %%CITATION = PHRVA,D68,103002;%%
  A.~Kusenko,
  %``Pulsar kicks from neutrino oscillations,''
  Int.\ J.\ Mod.\ Phys.\  D {\bf 13}, 2065 (2004).
%  [arXiv:astro-ph/0409521].
  %%CITATION = IMPAE,D13,2065;%%
  M.~Barkovich, J.C.D'Olivo and R.~Montemayor,
  %``Active-sterile neutrino oscillations and pulsar kicks,''
  Phys.\ Rev.\  D {\bf 70}, 043005 (2004).
%  [arXiv:hep-ph/0402259].
  %%CITATION = PHRVA,D70,043005;%%

\bibitem{Bode:2000gq}
  P.~Bode, J.P.~Ostriker and N.~Turok,
  %``Halo Formation in Warm Dark Matter Models,''
  Astrophys.\ J.\  {\bf 556}, 93 (2001).
%  [arXiv:astro-ph/0010389].
  %%CITATION = ASTRO-PH 0010389;%%
  V.~Avila-Reese, P.~Colin, O.~Valenzuela, E.~D'Onghia and
C.~Firmani,
%``Formation and structure of halos in a warm dark matter cosmology,''
Astrophys.\ J.\  {\bf 559}, 516 (2001).%
% [arXiv:astro-ph/0010525].
%%CITATION = ASTRO-PH 0010525;%%
  D.~Cumberbatch and J.~Silk,
  %``Accounting for the Unresolved X-ray Background with Sterile Neutrino Dark
  %Matter,''
  AIP Conf.\ Proc.\  {\bf 957} (2007) 375
  [arXiv:0709.0279 [astro-ph]].
  %%CITATION = APCPC,957,375;%%

\bibitem{Dalcanton:2000hn}
  J.J.~Dalcanton and C.J.~Hogan,
%``Halo Cores and Phase Space Densities: Observational Constraints on Dark
  %Matter Physics and Structure Formation,''
  Astrophys.\ J.\  {\bf 561}, 35 (2001).
%  [arXiv:astro-ph/0004381].
  F.C.~van den Bosch and R.A.~Swaters,
  %``Dwarf Galaxy Rotation Curves and the Core Problem of Dark Matter Halos,''
  Mon.\ Not.\ Roy.\ Astron.\ Soc.\  {\bf 325}, 1017 (2001).
% [arXiv:astro-ph/0006048].
  R.A.~Swaters, B.F.~Madore, F.C.V.~Bosch and M.~Balcells,
  %``The Central Mass Distribution in Dwarf and Low Surface Brightness
  %Galaxies,''
  Astrophys.\ J.\  {\bf 583}, 732 (2003).
%  [arXiv:astro-ph/0210152].
  G.~Gentile, P.~Salucci, U.~Klein, D.~Vergani and P.~Kalberla,
  %``The cored distribution of dark matter in spiral galaxies,''
  Mon.\ Not.\ Roy.\ Astron.\ Soc.\  {\bf 351}, 903 (2004).
% [arXiv:astro-ph/0403154].
  %%CITATION = MNRAA,351,903;%%
  J.D.~Simon, A.D.~Bolatto, A.~Leroy, L.~Blitz and E.L.~Gates,
  %``High-Resolution Measurements of the Halos of Four Dark Matter-Dominated
  %Galaxies: Deviations from a Universal Density Profile,''
  Astrophys.\ J.\  {\bf 621}, 757 (2005).
%  [arXiv:astro-ph/0412035].
  E.~Zackrisson, N.~Bergvall, T.~Marquart and G.~Ostlin,
  %``The dark matter halos of the bluest low surface brightness galaxies,''
  Astron.\ Astrophys.\ {\bf 452}, 857, (2006).
  %  arXiv:astro-ph/0603523.
  %% CITATION = ASTRO-PH/0603523;%%
  R.~Kuzio de Naray, S.S.~McGaugh, W.J.G.~de Blok and A.~Bosma,
  %``High Resolution Optical Velocity Fields of 11 Low Surface Brightness
  %Galaxies,''
  Astrophys.\ J.\ Suppl.\  {\bf 165}, 461 (2006).
%  [arXiv:astro-ph/0604576].
  %%CITATION = APJSA,165,461;%%
  T.~Goerdt, B.~Moore, J.I.~Read, J.~Stadel and M.~Zemp,
  %``Does the Fornax dwarf spheroidal have a central cusp or core?,''
  Mon.\ Not.\ Roy.\ Astron.\ Soc.\  {\bf 368}, 1073 (2006).
%  [arXiv:astro-ph/0601404].
  %%CITATION = ASTRO-PH 0601404;%%
  L.E.~Strigari, J.S.~Bullock, M.~Kaplinghat, A.V.~Kravtsov, O.Y.~Gnedin, K.~Abazajian and A.A.~Klypin,
  %``A Large Dark Matter Core in the Fornax Dwarf Spheroidal Galaxy?,''
  Astrophys.\ J.\  {\bf 652}, 306 (2006).
%  [arXiv:astro-ph/0603775].
  %%CITATION = ASTRO-PH 0603775;%%
  %%Cited 12 times in SPIRES-HEP


\bibitem{Kauffmann:1993gv}
  G.~Kauffmann, S.D.M.~White and B.~Guiderdoni,
  %``The Formation and Evolution of Galaxies Within Merging Dark Matter
  %Haloes,''
  Mon.\ Not.\ Roy.\ Astron.\ Soc.\  {\bf 264}, 201 (1993).
  %%CITATION = MNRAA,264,201;%%
  A.A.~Klypin, A.V.~Kravtsov, O.~Valenzuela and F.~Prada,
  %``Where are the missing galactic satellites?,''
  Astrophys.\ J.\  {\bf 522}, 82 (1999).
%  [arXiv:astro-ph/9901240].
  %%CITATION = ASTRO-PH 9901240;%%
  B.~Moore, S.~Ghigna, F.~Governato, G.~Lake, T.~Quinn, J.~Stadel and P.~Tozzi,
  %``Dark matter substructure within galactic halos,''
  Astrophys.\ J.\  {\bf 524}, L19 (1999).
  %%CITATION = ASJOA,524,L19;%%
  %  astro-ph/9907411

\bibitem{historical}
Sterile/active oscillations and BBN: D. Kirilova, Dubna preprint
JINR E2-88-301.
  R.~Barbieri and A.~Dolgov,
  %``BOUNDS ON STERILE-NEUTRINOS FROM NUCLEOSYNTHESIS,''
  Phys.\ Lett.\ B {\bf 237}, 440 (1990).
  %%CITATION = PHLTA,B237,440;%%
K.~Enqvist, K.~Kainulainen and J.~Maalampi,
  %``Resonant neutrino transitions and nucleosynthesis,''
  Phys.\ Lett.\ B {\bf 249} (1990) 531.
  %%CITATION = PHLTA,B249,531;%%
  K.~Kainulainen,
  %``Light Singlet Neutrinos And The Primordial Nucleosynthesis,''
  Phys.\ Lett.\ B {\bf 244}, 191 (1990).
  %%CITATION = PHLTA,B244,191;%%
  R.~Barbieri and A.~Dolgov,
  %``Neutrino oscillations in the early universe,''
  Nucl.\ Phys.\ B {\bf 349}, 743 (1991).
  %%CITATION = NUPHA,B349,743;%%
  K.~Enqvist, K.~Kainulainen and M.~J.~Thomson,
  %``Stringent cosmological bounds on inert neutrino mixing,''
  Nucl.\ Phys.\ B {\bf 373}, 498 (1992).
  %%CITATION = NUPHA,B373,498;%%
    J.~M.~Cline,
  % ``Constraints on almost Dirac neutrinos from neutrino - anti-neutrino
  %oscillations,''
  Phys.\ Rev.\ Lett.\  {\bf 68}, 3137 (1992).
  %%CITATION = PRLTA,68,3137;%%
X.~Shi, D.~N.~Schramm and B.~D.~Fields,
  %``Constraints on neutrino oscillations from big bang nucleosynthesis,''
  Phys.\ Rev.\ D {\bf 48}, 2563 (1993)
  [arXiv:astro-ph/9307027].
  %%CITATION = ASTRO-PH 9307027;%%
  E.~Lisi, S.~Sarkar and F.~L.~Villante,
  %``The big bang nucleosynthesis limit on N(nu),''
  Phys.\ Rev.\ D {\bf 59}, 123520 (1999)
  [arXiv:hep-ph/9901404].
  %%CITATION = HEP-PH 9901404;%%
K.~N.~Abazajian,
  %``Telling three from four neutrinos with cosmology,''
  Astropart.\ Phys.\  {\bf 19}, 303 (2003)
  [arXiv:astro-ph/0205238].
  %%CITATION = ASTRO-PH 0205238;%%
  D.~P.~Kirilova and M.~P.~Panayotova,
  %``Relaxed constraints on neutrino oscillation parameters,''
  JCAP {\bf 0612} (2006) 014
  [arXiv:astro-ph/0608103].
  %%CITATION = JCAPA,0612,014;%%
  C.~T.~Kishimoto, G.~M.~Fuller and C.~J.~Smith,
  %``Coherent active-sterile neutrino flavor transformation in the early
  %universe,''
  Phys.\ Rev.\ Lett.\  {\bf 97} (2006) 141301
  [arXiv:astro-ph/0607403].
  %%CITATION = PRLTA,97,141301;%%
  K.~Abazajian, N.~F.~Bell, G.~M.~Fuller and Y.~Y.~Y.~Wong,
  %``Cosmological lepton asymmetry, primordial nucleosynthesis, and sterile
  %neutrinos,''
  Phys.\ Rev.\  D {\bf 72} (2005) 063004
  [arXiv:astro-ph/0410175].
  %%CITATION = PHRVA,D72,063004;%%
  Y.~Z.~Chu and M.~Cirelli,
  %``Sterile neutrinos, lepton asymmetries, primordial elements: How much of
  %each?,''
  Phys.\ Rev.\  D {\bf 74} (2006) 085015
  [arXiv:astro-ph/0608206].
  %%CITATION = PHRVA,D74,085015;%%
  C.~J.~Smith, G.~M.~Fuller, C.~T.~Kishimoto and K.~N.~Abazajian,
  %``Light element signatures of sterile neutrinos and cosmological lepton
  %numbers,''
  Phys.\ Rev.\  D {\bf 74} (2006) 085008
  [arXiv:astro-ph/0608377].
  %%CITATION = PHRVA,D74,085008;%%
    K.~Abazajian,
  %``Production and evolution of perturbations of sterile neutrino dark
  %matter,''
  Phys.\ Rev.\  D {\bf 73} (2006) 063506
  [arXiv:astro-ph/0511630].
  %%CITATION = PHRVA,D73,063506;%%


\bibitem{DiBari:2001ua}
  P.~Di Bari,
  %``Update on neutrino mixing in the early universe,''
  Phys.\ Rev.\  D {\bf 65} (2002) 043509
  [Addendum-ibid.\  D {\bf 67} (2003) 127301]
  [arXiv:hep-ph/0108182].
  %%CITATION = PHRVA,D65,043509;%%
    P.~Di Bari,
  %``Addendum to: Update on neutrino mixing in the early universe,''
  Phys.\ Rev.\  D {\bf 67} (2003) 127301
  [arXiv:astro-ph/0302433].
  %%CITATION = PHRVA,D67,127301;%%
  R.~H.~Cyburt, B.~D.~Fields and K.~A.~Olive,
  %``Primordial Nucleosynthesis in Light of WMAP,''
  Phys.\ Lett.\  B {\bf 567} (2003) 227
  [arXiv:astro-ph/0302431].
  %%CITATION = PHLTA,B567,227;%%
    V.~Barger, J.~P.~Kneller, H.~S.~Lee, D.~Marfatia and G.~Steigman,
  %``Effective number of neutrinos and baryon asymmetry from BBN and WMAP,''
  Phys.\ Lett.\  B {\bf 566} (2003) 8
  [arXiv:hep-ph/0305075].
  %%CITATION = PHLTA,B566,8;%%
  G.~Mangano, G.~Miele, S.~Pastor, T.~Pinto, O.~Pisanti and P.~D.~Serpico,
  %``Relic neutrino decoupling including flavour oscillations,''
  Nucl.\ Phys.\  B {\bf 729} (2005) 221
  [arXiv:hep-ph/0506164].
  %%CITATION = NUPHA,B729,221;%%

\bibitem{Hannestad:2005ex}
  S.~Hannestad and G.~Raffelt,
  %``Constraining invisible neutrino decays with the cosmic microwave
  %background,''
  Phys.\ Rev.\  D {\bf 72} (2005) 103514
  [arXiv:hep-ph/0509278].
  %%CITATION = PHRVA,D72,103514;%%
  M.~Cirelli and A.~Strumia,
  %``Cosmology of neutrinos and extra light particles after WMAP3,''
  JCAP {\bf 0612} (2006) 013
  [arXiv:astro-ph/0607086].
  %%CITATION = JCAPA,0612,013;%%

\bibitem{Bell:2005dr}
  N.~F.~Bell, E.~Pierpaoli and K.~Sigurdson,
  %``Cosmological signatures of interacting neutrinos,''
  Phys.\ Rev.\  D {\bf 73} (2006) 063523
  [arXiv:astro-ph/0511410].
  %%CITATION = PHRVA,D73,063523;%%

\bibitem{Hannestad:2004qu}
  S.~Hannestad,
  %``Structure formation with strongly interacting neutrinos: Implications  for
  %the cosmological neutrino mass bound,''
  JCAP {\bf 0502} (2005) 011
  [arXiv:astro-ph/0411475].
  %%CITATION = JCAPA,0502,011;%%

\bibitem{Friedland:2006ta}
  A.~Friedland and A.~Gruzinov,
  %``Neutrino signatures of supernova turbulence,''
  arXiv:astro-ph/0607244.
  %%CITATION = ASTRO-PH/0607244;%%
  S.~Choubey, N.~P.~Harries and G.~G.~Ross,
  %``Turbulent supernova shock waves and the sterile neutrino signature in
  %megaton water detectors,''
  Phys.\ Rev.\  D {\bf 76} (2007) 073013
  [arXiv:hep-ph/0703092].
  %%CITATION = PHRVA,D76,073013;%%


\bibitem{Hinshaw:2006ia}
  G.~Hinshaw {\it et al.}  [WMAP Collaboration],
  %``Three-year Wilkinson Microwave Anisotropy Probe (WMAP) observations:
  %Temperature analysis,''
  Astrophys.\ J.\ Suppl.\  {\bf 170} (2007) 288
  [arXiv:astro-ph/0603451].
  %%CITATION = APJSA,170,288;%%
  L.~Page {\it et al.}  [WMAP Collaboration],
  %``Three year Wilkinson Microwave Anisotropy Probe (WMAP) observations:
  %Polarization analysis,''
  Astrophys.\ J.\ Suppl.\  {\bf 170} (2007) 335
  [arXiv:astro-ph/0603450].
  %%CITATION = APJSA,170,335;%%
  D.~N.~Spergel {\it et al.}  [WMAP Collaboration],
  %``Wilkinson Microwave Anisotropy Probe (WMAP) three year results:
  %Implications for cosmology,''
  Astrophys.\ J.\ Suppl.\  {\bf 170} (2007) 377
  [arXiv:astro-ph/0603449].
  %%CITATION = APJSA,170,377;%%

\bibitem{Tegmark:2003uf}
  M.~Tegmark {\it et al.}  [SDSS Collaboration],
  %``The 3D power spectrum of galaxies from the SDSS,''
  Astrophys.\ J.\  {\bf 606} (2004) 702
  [arXiv:astro-ph/0310725].
  %%CITATION = ASJOA,606,702;%%

\bibitem{Peiris:2006sj}
  H.~Peiris and R.~Easther,
  %``Slow roll reconstruction: Constraints on inflation from the 3 year WMAP
  %dataset,''
  JCAP {\bf 0610} (2006) 017
  [arXiv:astro-ph/0609003].
  %%CITATION = JCAPA,0610,017;%%

\bibitem{Guth:1980zm}
  A.~H.~Guth,
  %``The Inflationary Universe: A Possible Solution To The Horizon And Flatness
  %Problems,''
  Phys.\ Rev.\  D {\bf 23} (1981) 347.
  %%CITATION = PHRVA,D23,347;%%
  A.~D.~Linde,
  %``A New Inflationary Universe Scenario: A Possible Solution Of The Horizon,
  %Flatness, Homogeneity, Isotropy And Primordial Monopole Problems,''
  Phys.\ Lett.\  B {\bf 108} (1982) 389.
  %%CITATION = PHLTA,B108,389;%%
  A.~Albrecht and P.~J.~Steinhardt,
  %``Cosmology For Grand Unified Theories With Radiatively Induced Symmetry
  %Breaking,''
  Phys.\ Rev.\ Lett.\  {\bf 48} (1982) 1220.
  %%CITATION = PRLTA,48,1220;%%

\bibitem{Cole:2005sx}
  S.~Cole {\it et al.}  [The 2dFGRS Collaboration],
  %``The 2dF Galaxy Redshift Survey: Power-spectrum analysis of the final
  %dataset and cosmological implications,''
  Mon.\ Not.\ Roy.\ Astron.\ Soc.\  {\bf 362} (2005) 505
  [arXiv:astro-ph/0501174].
  %%CITATION = MNRAA,362,505;%%

%
\bibitem{Heckler:1993nc}
  A.~Heckler and C.~J.~Hogan,
  %``Neutrino heat conduction and inhomogeneities in the early universe,''
  Phys.\ Rev.\  D {\bf 47} (1993) 4256.
  %%CITATION = PHRVA,D47,4256;%%

%
\bibitem{Schwarz:2003du}
  D.~J.~Schwarz,
  %``The first second of the universe,''
  Annalen Phys.\  {\bf 12} (2003) 220
  [arXiv:astro-ph/0303574].
  %%CITATION = ANPYA,12,220;%%

\bibitem{Schmid:1996qd}
  C.~Schmid, D.~J.~Schwarz and P.~Widerin,
  %``Peaks above the Harrison-Zel'dovich spectrum due to the quark-gluon to
  %hadron transition,''
  Phys.\ Rev.\ Lett.\  {\bf 78} (1997) 791
  [arXiv:astro-ph/9606125].
  %%CITATION = PRLTA,78,791;%%
  C.~Schmid, D.~J.~Schwarz and P.~Widerin,
  %``Amplification of cosmological inhomogeneities from the QCD transition,''
  Phys.\ Rev.\  D {\bf 59} (1999) 043517
  [arXiv:astro-ph/9807257].
  %%CITATION = PHRVA,D59,043517;%%
  P.~Widerin and C.~Schmid,
  %``Primordial black holes from the QCD transition?,''
  arXiv:astro-ph/9808142.
  %%CITATION = ASTRO-PH/9808142;%%

\end{thebibliography}
\end{document}